\documentclass{aa}
\usepackage{txfonts}
\usepackage{graphicx}
\usepackage{natbib}
\bibpunct{(}{)}{;}{a}{}{,} 
\usepackage{color}

\begin{document}

\title{Chromospheric heating and generation of plasma outflows by impulsively generated two-fluid magnetoacoustic waves}

\authorrunning{R.~Niedziela et al.}
\titlerunning{Magnetoacoustic waves in the solar atmosphere}

   \author{R. Niedziela\inst{1}
          \and
          K. Murawski\inst{1}
          \and 
          S. Poedts\inst{2,1}
          }

   \institute{Institute of Physics, University of M.\ Curie-Sk{\l}odowska, 
              Pl.\ M.\ Curie-Sk{\l}odowskiej 1, 20-031 Lublin, Poland\\
         \and
            Centre for Mathematical Plasma Astrophysics / Department of Mathematics, KU Leuven, Celestijnenlaan 200B, 3001 Leuven, Belgium\\
             }



\abstract
   {The origin of the heating of the solar atmosphere is still an unsolved problem.
   As the photosphere and chromosphere radiate more energy than the solar corona, it is challenging but important to reveal all the mechanisms that contribute to plasma heating there. 
   Ion--neutral collisions could play an important role.}
   {We aim to investigate the impulsively generated two-fluid magnetoacoustic waves in the partially ionized solar chromosphere and to study the associated heating and plasma outflows, which higher up may result in nascent solar wind. }
   {To describe the plasma dynamics, we applied a two-fluid model in which ions+electrons and neutrals are treated as separate fluids.
   We solved the two-fluid equations numerically using the JOANNA code. }
   {We show that magnetoacoustic waves triggered in the photosphere by localised velocity pulses can steepen into shocks which heat the chromosphere through ion--neutral collisions.
    Pulses of greater amplitude heat plasma more effectively and generate larger plasma outflows.
   Rising the altitude at which the pulse is launched results in opposite effects, mainly in local cooling of the chromosphere and slower plasma outflows.}
  {Even a solitary pulse results in a train of waves. These waves can transform into shock waves and release thermal energy, heating the chromosphere significantly.
  A pulse can drive vertical flows which higher up can result in the origin of the solar wind.}
\keywords{Sun: activity - Sun: chromosphere}

\maketitle

\section{Introduction}
One of the main problems of heliophysics concerns the source of the thermal energy required to heat the solar atmosphere,  the lowest layer of which is about $ 500 \; \rm{km}$ thick and is called the photosphere.
The chromosphere, which is about $1.5\;\rm{Mm}$ thick, caps this layer. A narrow  transition region separates the chromosphere from the solar corona above it.
Temperatures in the photosphere vary from about $5600\;\rm{K}$ at its bottom to about $4300\;\rm{K}$ at its top. The temperature of the middle of the chromosphere rises to  about $10^4\;\rm{K}$   ($\pm 800-\pm 1800\;$km above the solar surface). Because of the temperatures in the photosphere and low and middle chromosphere,  the plasma in these layers is only partially ionized.
Recombination, photoionization, and chemical ionization were discussed by \cite{2002ApJ...572..626C}.
At the top of the photosphere, the ionization degree is as low as $10^{-4}$, and so there are about $10^4$ neutrals per ion. In the chromosphere, the ratio of ions 
to neutrals grows with height, and therefore with\ temperature \citep{2014PhPl21i2901K}.
As a result of its $1 - 3\;\rm{MK}$ average temperature, the plasma in the corona is almost fully ionized.

The plasma density is also higher in the chromosphere and thus it radiates more than the corona. Hence, an additional mechanism of heating is required to compensate for these radiative losses \citep{narain1996chromospheric}.
One of the sources of this heating may be associated with collisions between particles, which lead to wave damping \citep{refId0}.
The idea of heating the solar chromosphere by acoustic waves is very old and was first developed by \cite{biermann1946deutung} and \cite{schwarzschild1948noise}, who suggested that acoustic waves play a major role in the heating of this atmospheric layer.
Later, the problem was widely studied, for example by \cite{1995ApJ440L29C}, \cite{2003ASPC286363U}, and \cite{2019ApJ87881K}. The latter authors focused on the acoustic waves with wave periods within the range $30-200\;\rm{s}$ and showed that these waves are able to heat the chromosphere.
The idea of heating the solar atmosphere without shock waves was studied by \cite{2020A&A...635A..28W} who showed that the thermal energy release ratio during ion--neutral collisions is sufficient to balance the radiative and thermal energy losses.
Recently, \cite{2021arXiv210509882K} focused on the studies of ion magnetoacoustic and neutral acoustic waves in a magnetic arcade and found that, in the chromosphere, the wave characteristics strongly depend on the magnetic field configuration.
The advantages of using the two-fluid model over the single-fluid model were studied by \citet{2011A&A529A82Z}.
The single-fluid (MHD) model is sufficiently accurate for slow processes but fails on timescales shorter than the ion--neutral collision time. Hence,  the two-fluid approximation should be used for studying faster processes.

The problem of atmospheric heating was also studied in the context of Alfv\'en waves \citep{101093mnras1163314, 1961ApJ134347O}.
The mechanisms by which these waves are dampened by ion--neutral collision were studied by \cite{De_Pontieu_2001}.
In the lower chromosphere, collisional effects were found to be the most important component of the dissipation mechanism \citep{refId03}.
As a matter of fact, Alfv\'en waves can dissipate their energy in the chromosphere as a result of ion--neutral collisions \citep{2005A&A4421091L, goodman2011conditions, articleSong, 2013A&A549A113Z, 2016ApJ81794A, 2016AGUFMSH21E2565S, 2017ApJ84020S}.
However, this mechanism is  only effective in the chromosphere and not in the photosphere or in the corona because of weak magnetization and strong ionization, respectively.
In other studies, \citet{2013A&A549A113Z} and \citet{2017ApJ84020S} focused on the dissipation of torsional Alfv\'en waves and showed that they can by effectively dissipated in the chromosphere. 

The mechanisms considered in the present paper for solar coronal heating are essentially based on magnetic field reconnection and resonant dissipation and/or phase-mixing of magnetohydrodynamic (MHD) waves. 
The latter mechanism was reviewed recently by \cite{2020SSRv..216..140V},
who showed that various wave modes may contribute to 
the heating of the solar corona. 
The effect of the cooling and heating on the dynamics of the MHD waves was investigated by \cite{2017ApJ...849...62N} who concluded that the perturbations of the thermal equilibrium, for example by\ slow magnetoacoustic waves, lead to a heating--cooling unbalance which acts as an energy exchange mechanism.
For more recent studies of thermal unbalance, see also \cite{2021SoPh..296...20P} who studied the damping of slow magnetosonic waves based on a new dispersion relation.
The problem of damping slow magnetosonic waves in the solar corona was also investigated by \cite{2021A&A...646A.155D},  considering nonzero $\beta$ plasma, and \cite{2019A&A...628A.133K} who presented three potential regimes of wave evolution depending on characteristic  timescales of the thermal unbalance.
Recently, \cite{2020A&A...644A..33K} constrained the coronal heating function by observations of slow magnetoacoustic waves.

The solar corona extends smoothly into the solar wind.
Studies of the solar wind were initiated by \citet{1951ZA29274B} and \citet{1965SSRv4666P}.
\citet{1958ApJ128664P} postulated that the interplanetary space is not void but that it is filled by the solar wind.
An early model of the solar wind with the power spectrum of Alfv\'en and kink waves was developed by \citet{1987SoPh109149T}.
Later on, a lot of attention was paid to the study of the formation of the solar wind by plasma outflows.
For instance, \citet{2005Sci308519T} found these outflows in coronal funnels at altitudes between 5 and $20\;$Mm above the photosphere.
Later, the problem was reconsidered by \citet{2019ApJ884127W} who proposed that jets and associated plasma outflows generated by the solar granulation may result in the formation of the solar wind higher up.
Moving upwardly, the plasma requires a source of momentum which can be provided by MHD waves \citep{2005SSRv12067O, 2006LRSP31M, 2007Sci3181574D}.
Alfv\'en waves, for instance, may drive plasma outflows \citep{1982SoPh7535H, 1999ApJ514493K, 2012ApJ7498M}.
A model for coronal heating and solar wind acceleration by low-frequency Alfv\'en wave turbulence was presented by \citet{2014ApJ78281V}.

The goal of the present paper is to focus on magnetoacoustic waves in the context of heating the chromosphere and the associated plasma outflows, and therefore we leave the role of Alfv\'en waves for future studies.
The novelty of this work is that we consider impulsively generated two-fluid magnetoacoustic waves in the context of solar  chromospheric heating and plasma outflows.
This paper is organized as follows. In Sect.~2, we present the two-fluid equations that are considered for describing the dynamics of the partially ionized solar atmosphere.  Numerical results are presented in Sect.~3. In Sect.~4 we conclude this paper by summarizing the main results and briefly discussing them.
%
\section{Two-fluid numerical model of the partially ionized solar atmosphere}
%
\subsection{Two-fluid equations}
To describe the low layers of the solar atmosphere we use two-fluid equations for ions + electrons ($_{i,e}$) and neutrals ($_n$) treated as separate interacting fluids \citep{Zaqarashvilietal2011,2014SSRv..184..107L,Oliveretal2016,Manevaetal2017,PopescuBraileanuetal2019}:
\begin{eqnarray}
\frac{\partial \varrho_{\rm i}}
{\partial t}+\nabla\cdot(\varrho_{\rm i} \mathbf{V}_{\rm i}) = 0
\, , 
\label{eq:ion_continuity} \\
\frac{\partial \varrho_{\rm n}}{\partial t}+
\nabla\cdot(\varrho_{\rm n} \mathbf{V}_{\rm n}) = 0 
\, , 
\label{eq:neutral_continuity} \\
\frac{\partial  (\varrho_{\rm i} \mathbf{V}_{\rm i})}{\partial t}
+
\nabla \cdot (\varrho_{\rm i} \mathbf{V}_{\rm i} \mathbf{V}_{\rm i}
+
p_{\rm ie} \mathbf{I}) 
= \\
\nonumber
\varrho_{\rm i} \mathbf{g} 
+ 
\frac{1}{\mu}(\nabla \times \mathbf{B}) \times \mathbf{B} 
+ 
\mathbf{S_{\rm m}}
\, ,
\label{eq:ion_momentum}\\
\frac{\partial (\varrho_{\rm n} \mathbf{V}_{\rm n})}{\partial t} 
+
\nabla \cdot (\varrho_{\rm n} \mathbf{V}_{\rm n} \mathbf{V}_{\rm n}
+
p_{\rm n} \mathbf{I})  
=
\varrho_{\rm n} \mathbf{g} 
- 
\mathbf{S_{\rm m}}
\, ,
\label{eq:neutral_momentum} 
\\
\frac{\partial E_{\rm i}}{\partial t}+\nabla\cdot\left[\left(E_{\rm i}+p_{\rm ie} + 
\frac{\mathbf{B}^2}{2\mu} \right)\mathbf{V}_{\rm i}-\frac{\mathbf{B}}{\mu}(\mathbf{V}_{\rm i}\cdot \mathbf{B}) 
\right]  
= \\
\nonumber
(\varrho_{\rm i} \mathbf{g}+\mathbf{S_{\rm m}})  
\cdot \mathbf{V}_{\rm i}
+ Q_{\rm i}
\, , 
\label{eq:ion_energy} 
\\
\frac{\partial E_{\rm n}}{\partial t}+\nabla\cdot[(E_{\rm n}+p_{\rm n})\mathbf{V}_{\rm n}] 
=
(\varrho_{\rm n} \mathbf{g} - \mathbf{S_{\rm m}})
\cdot \mathbf{V}_{\rm n} 
+ Q_{\rm n} 
\, ,
\label{eq:neutral_energy} \\
\frac{\partial \mathbf{B}}{\partial t}
=
\nabla \times (\mathbf{V_{\rm i} \times }\mathbf{B})
\, ,
\hspace{3mm}
\quad
 \nabla \cdot{\mathbf B} = 0
 \, ,
\label{eq:ions_induction} \\
E_{\rm i} = \frac{\varrho_{\rm i}\mathbf{V}_{\rm i}^2}{2} + 
\frac{p_{\rm ie}}{\gamma -1 } +\frac{{\mathbf B}^2}{2\mu}
\, , 
\hspace{3mm}
E_{\rm n} = \frac{\varrho_{\rm n}\mathbf{V}_{\rm n}^2}{2} + 
\frac{p_{\rm n}}{\gamma -1 }
\, .
\end{eqnarray}
Here, $\textbf{I}$ is the identity matrix, $p_{ie,n}$ denote the ion + electron and neutral gas pressures, while $\varrho_{i,n}$ denote ion and neutral mass densities, and $\textbf{V}_{i,n}$ the ion and neutral velocities. Moreover, $\textbf{g} = [0,-g,0]$ is gravitational acceleration with its magnitude $g=274.78 \; \rm{m \: s}^{-2}$, $\textbf{B}$ is the magnetic field pointing in the $y-$direction, and $\mu$ the magnetic permeability of the medium. In Eqs. (3) - (6), $\textbf{S}_m$ and $Q_{i,n}$ correspond to the collisional momentum and energy exchange terms, respectively. They are given as \citep{Meier2012,Oliveretal2016}
\begin{eqnarray}
\mathbf{S_{\rm m}} &=& 
\varrho_i 
\nu_{\rm in}
(\mathbf{V_{\rm n}} - \mathbf{V_{\rm i}})
\, ,
\\
Q_{\rm i} &=&
 \frac{1}{2}\nu_{\rm in}\varrho_{\rm i}
(\mathbf{V_{\rm i}}-\mathbf{V_{\rm n}})^2 
+
\frac{k_{\rm B}\nu_{\rm in}\varrho_{\rm i}}{(\gamma-1)m_{\rm n}}
(T_{\rm n}-T_{\rm i})
\, ,
\\
Q_{\rm n} &=&
 \frac{1}{2}\nu_{\rm in}\varrho_{\rm i}
(\mathbf{V_{\rm i}}-\mathbf{V_{\rm n}})^2 
+
\frac{k_{\rm B}\nu_{\rm in}\varrho_{\rm i}}{(\gamma-1)m_{\rm n}}
(T_{\rm i}-T_{\rm n}) \, . 
\end{eqnarray}
The ion--neutral  collision  frequency, $\nu_{in}$, is defined as \citep{Braginskii1965,Ballesteretal2018}
\begin{eqnarray}
 \nu_{\rm in} = 
  \frac{4}{3} 
 \frac{\sigma_{\rm in}\varrho_{\rm n}}{m_{\rm H}(\mu_{\rm i}+\mu_{\rm n})}
 \sqrt{ \frac{8k_{\rm B}}{\pi m_{\rm H}} 
\left(
\frac{T_{\rm i}}{\mu_{\rm i}}+\frac{T_{\rm n}}{\mu_{\rm n}}
\right) }
      \, .
\end{eqnarray}
In the above equation, $m_H$ is the hydrogen mass, $\sigma_{\rm in} = 1.4 \cdot 10^{-15} \; \rm{cm^2}$ is the collision cross-section \citep{2013A&A...554A..22V}, $\mu_{\rm i}=0.58$ and $\mu_{\rm n}=1.21$ are the mean masses of ions and neutrals, respectively, and $T_{i,n}$ are the temperatures specified by the ideal gas laws,
\begin{equation}
p_{\rm ie}=\frac{k_{\rm B}}{m_i}\varrho_{\rm i}T_{\rm i}\, , \quad\hbox{and}\quad
p_{\rm n} =\frac{k_{\rm B}}{m_n}\varrho_{\rm n}T_{\rm n}\, .
\label{eq:pressures}
\end{equation}
Here, $m_{\rm i} = m_{\rm H} \mu_{\rm i}$, and
$m_{\rm n} = m_{\rm H} \mu_{\rm n}$, 
where $m_H$ corresponds to the mass of a hydrogen atom, $k_B$ is the Boltzmann constant, and $\gamma = 1.4$ corresponds to the specific heat ratio.
As the photospheric plasma contains bi-atomic 
molecules, we decided to consider $\gamma=1.4$ instead of $5/3$.
The latter value is suitable for plasma that only consists of atoms.
For simplicity, we neglected thermal conduction and other nonadiabatic and nonideal terms in the two-fluid equations.
We should expect that thermal conduction will impact wave propagation by diffusing thermal energy into less localized and thus less intensively plasma-heated regions.

The two-fluid equations do not describe the top of the convection zone and the photosphere  very well, which are included in the simulation domain, but they do accurately describe the chromosphere and the low corona, which are the focus of our simulation model. Problems with the two-fluid equations result from comparable values of characteristic gyro- and collision frequencies in the photosphere \cite[Fig.~1, top]{2014PhPl...21i2901K}, while fluid models require that the collision frequencies be much higher than the gyro frequencies.
\subsection{Magnetohydrostatic equilibrium}
As an initial condition, we assume that the atmosphere is static ($\textbf{V}_{i,n}= \textbf{0}$) and is in hydrostatic equilibrium with the ion and neutral pressure gradients being balanced by the gravity forces on both species:
\begin{equation}
    - \nabla p_{i,n} + \varrho_{i,n} \textbf{g} = \textbf{0} \, .
\end{equation}

Using the ideal gas laws in Eq.~(13) and the $y$-components of Eq.~(14), we express the background gas pressures and mass densities as
\begin{equation}
    p_{i,n}(y) = p_{i,n 0} \, \exp{\left(- \int^y_{y_r} \frac{dy}{\Lambda_{i,n}(y)} \right)} \:,
\end{equation}
\begin{equation}
    \varrho_{i,n}(y) = \frac{p_{i,n}(y)}{g \Lambda_{i,n}(y)} \, ,
\end{equation}
where $\Lambda_{i,n} = k_B T / (\mu_{i,n} g)$ are ion and neutral pressure scale heights, respectively, while $p_{i0}=0.01 \; \rm{Pa}$ and $p_{n0}=0.0003 \; \rm{Pa}$ denote the corresponding gas pressures at the reference level, taken here at $y= y_r = 50 \; \rm{Mm}$. \\
%
%
\begin{figure}[ht]
\begin{center}
\hspace{-0.2cm}
\includegraphics[width=0.45\textwidth]{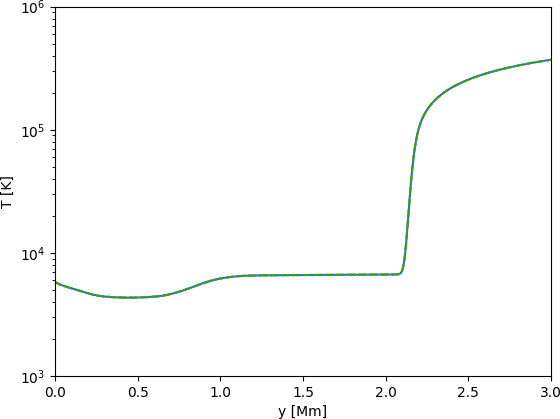}
\hspace{0.5cm}
\includegraphics[width=0.45\textwidth]{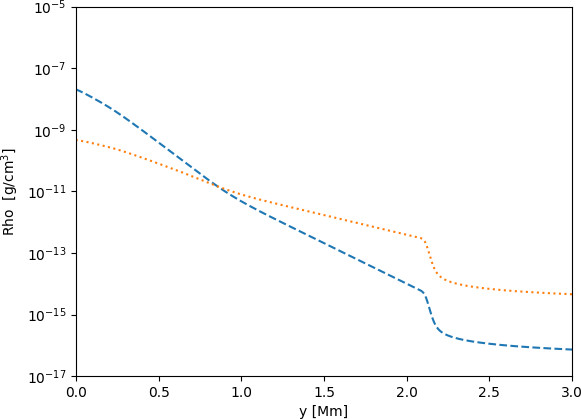}
\vspace{0.25cm}
\includegraphics[width=0.45\textwidth]{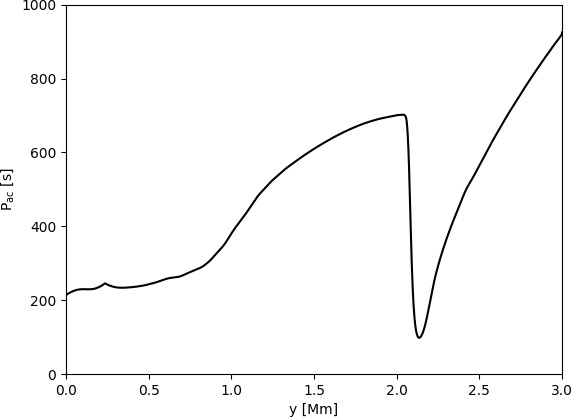}
\end{center}
\vspace{-0.25cm}
\caption{Variation with height of the equilibrium temperature (top), mass densities (middle) of neutrals (dashed line) and ions (dotted line), and acoustic cut-off period (bottom).}
\label{fig1}
\end{figure}
%
%
We consider the height-dependent temperature, $T(y),$ model of \citet{AvrettLoeser2008}; see Fig.~1 (top). This model also determines the equilibrium mass densities and gas pressure profiles.
As the electron mass is very small in comparison to the mass of neutrals and ions, we neglect the electron dynamics.  
We note that the ion mass density is lower than that of the neutral component in the photosphere and low chromosphere (Fig.~1, middle).
However, above the altitude of $y \approx 0.9 \; \rm{Mm}$ and especially above the transition region, which is located at $y \approx 2.1 \; \rm{Mm}$, neutrals are much less numerous than ions.
The presence of gravity in the system leads to the appearance of an acoustic cut-off period (Fig.~1, bottom).
The formula is given as
\begin{equation}
P_{\rm ac} = \frac{4 \pi \Lambda}{ C_S \sqrt{1 + 2 \frac{d \Lambda}{dy}} },
\end{equation}
where $\Lambda$ is the pressure scale height, and $C_s = \sqrt{\gamma (p_i + p_n)/(\varrho_i + \varrho_n)}$ is the sound speed. 
\\
The pressure scale height $\Lambda$ is responsible for the propagation of the waves when their period $P$ is smaller than the $P_{\rm ac}$. Otherwise, if this value is greater than or equal to $P_{\rm ac}$, the waves are evanescent \citep{Lamb}.
It is worth noting that the $P_{\rm ac}$  considered here is a local quantity that varies with height \citep{2006PhRvE..73c6612M}.
\\
The hydrostatic equilibrium  described  above is overlaid by a uniform vertical magnetic field of magnitude $B_y = 10 \; \rm{Gs}$.
Such a magnetic field is current-free ($\nabla \times \textbf{B} / \mu = 0$) and therefore force-free $\big((\nabla \times \textbf{B}) \times \textbf{B} / \mu = 0 \big)$, and thus does not alter the hydrostatic equilibrium.
\subsection{Impulsive perturbations}
At $t=0 \; \rm{s}$, when the numerical simulations are started, we perturb the magnetohydrostatic equilibrium with a localized signal in the vertical components of the ion and neutral velocities:\begin{equation}
V_{\rm iy}(y,t=0)=V_{ny}(y,t=0)=A \, \exp\left(-\left(\frac{y-y_{\rm 0}}{w}\right)^2\right) \, ,
\end{equation}
where $A$ is the amplitude of the pulse, $w$ its width, and $y_{\rm 0}$ the height from which the pulse is launched.
We consider the following cases: $A=0.5 \; \rm{km \: s^{-1}}$, $A=1 \; \rm{km \: s^{-1}}$, $A=2 \; \rm{km \: s^{-1}}$, and $A=-5 \; \rm{km \: s^{-1}}$, combined with different widths of the pulse, namely $w=0.1 \; \rm{Mm}$, $w=0.25 \; \rm{Mm}$, and $w=0.3 \; \rm{Mm}$; and various launching heights, namely\ $y_{\rm 0} = 0 \; \rm{Mm}$, $y_{\rm 0} = 0.25 \; \rm{Mm}$, and $y_{\rm 0} = 0.5 \; \rm{Mm}$.
The first two values of the amplitude $A$ represent a typical flow associated with the solar granulation \citep[e.g.,] [] {refId01}, while the third value corresponds to a stronger signal, and the last value mimics downdrafts.
The implied maximum value of the pulse amplitude $A=0.5 \; \rm{km} \; \rm{s}^{-1}$ is about six times larger  than the local Alfv\'en speed, which for the value of $B_y = 10 \; \rm{Gs}$ used here is about $C_A=0.08 \; \rm{km} \; \rm{s}^{-1}$ at $y = 0 \; \rm{Mm}$ and much smaller than the local sound speed which is about $C_S=7.5 \; \rm{km} \; \rm{s}^{-1}$ at $y = 0 \; \rm{Mm}$.
The characteristic width of the pulse, $w = 300 \; \rm km$, is comparable to the size of the smallest granules.
The difference between the positive and negative values of the amplitude results in the ion and neutral fluids being pushed upwards and downwards, respectively. However, both negative and positive pulses spread into upward and downward propagating waves.
We note that $y_{\rm 0} = 0 \; \rm{Mm}$ corresponds to the bottom of the photosphere, $y_{\rm 0} = 0.25 \; \rm{Mm}$ to the middle of the photosphere, and $y_{\rm 0} = 0.5 \; \rm{Mm}$ is associated with its top.
For simplicity, we consider only pulses in the ion and neutral vertical velocity components.
The physical nature of realistic perturbations is indeed more complex than the dependence presented by Eq.~(18).
Moreover, we consider an impulsive perturbation launched at $t = 0 \; \rm{s}$, without first relaxing the system, because the relative numerical errors are small in all atmospheric layers (Fig.~2).
%
\section{Numerical results}
The one-dimensional numerical simulations are performed with the JOANNA code \citep{Wojciketal2019a}, which solves the above-mentioned two-fluid equations.
We assume that the system has a slab geometry and that it is invariant along the $x$- and $z$-directions ($\partial / \partial x = \partial / \partial z =0$), and the $z-$components of the velocities and the magnetic field are both set to zero ($V_{iz} = V_{nz} = B_z = 0$).
Applying these simplifications, we remove Alfv\'en waves from the system which guides (affected by gravity) magnetoacoustic waves.
Internal-gravity waves are absent in the system as they are unable to propagate along the gravity action.\\
The simulation domain is specified as $- 0.5 \leq y \leq 60.0\;$Mm. 
The distance below $6.87\;$Mm is divided into $1024$ cells, and therefore the size of the grid cells is $\Delta y \approx 7.2 \; \rm{km}$. Higher up, the grid is stretched and the remainder of the domain is covered by $128$ cells.
At the top and bottom boundaries of the simulation region, we set and hold all plasma quantities 
fixed to their equilibrium values, given by Eqs.~(15) and (16).
%
%
\begin{figure}
\begin{center}
\hspace{-0.2cm}
\includegraphics[width=0.48\textwidth]{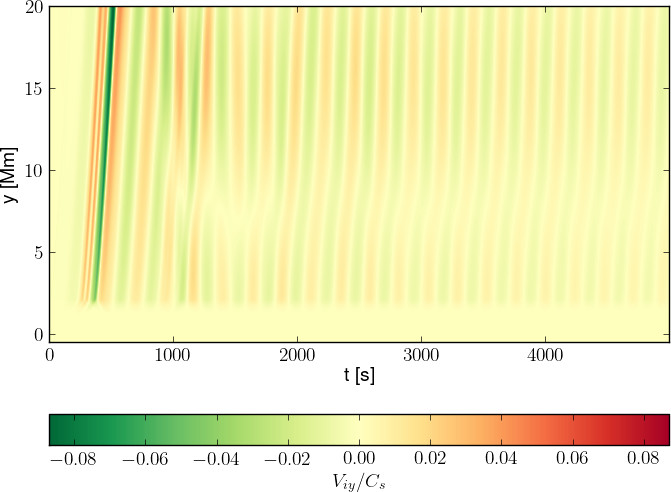}
\hspace{0.5cm}
\includegraphics[width=0.48\textwidth]{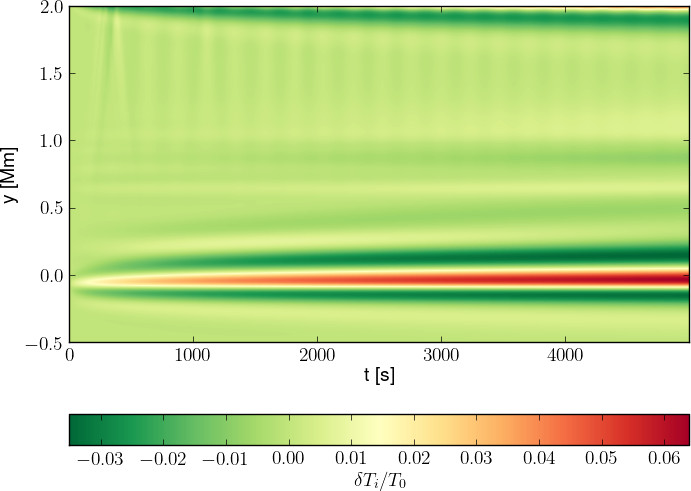}
\end{center}
\vspace{-0.25cm}
\caption{Time--distance plots for $V_{\rm iy}/C_s$ (top) and $\delta T_i / T$ (bottom) in the case of no pulse ($A=0$).}
\label{fig2}
\end{figure}
%
%
%
\subsection{Numerical test}
%
%
\begin{figure*}[ht]
    \begin{center}
    \hspace{-0.2cm}
    \includegraphics[width=0.48\textwidth]{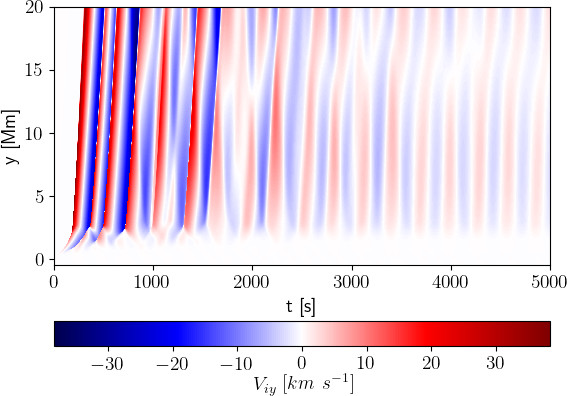}
    \hspace{0.5cm}
    \vspace{0.25cm}
    \includegraphics[width=0.48\textwidth]{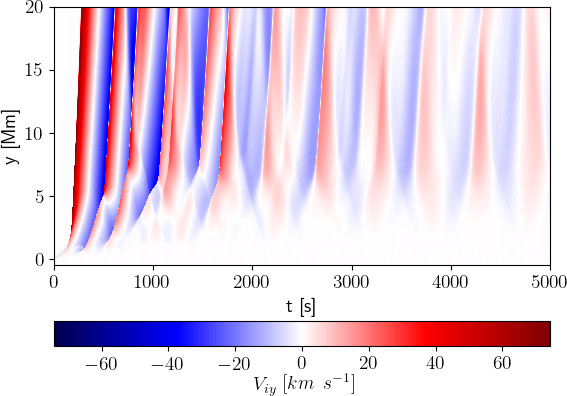}
    \vspace{0.25cm}
    \includegraphics[width=0.48\textwidth]{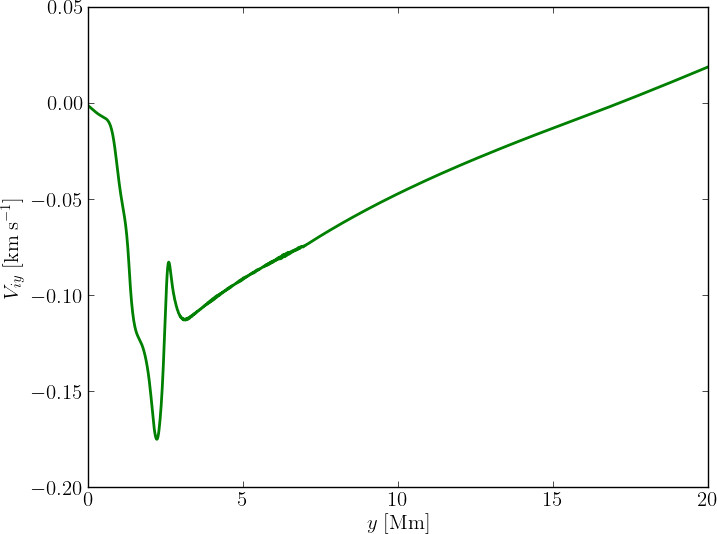}
    \hspace{0.5cm}
    \includegraphics[width=0.48\textwidth]{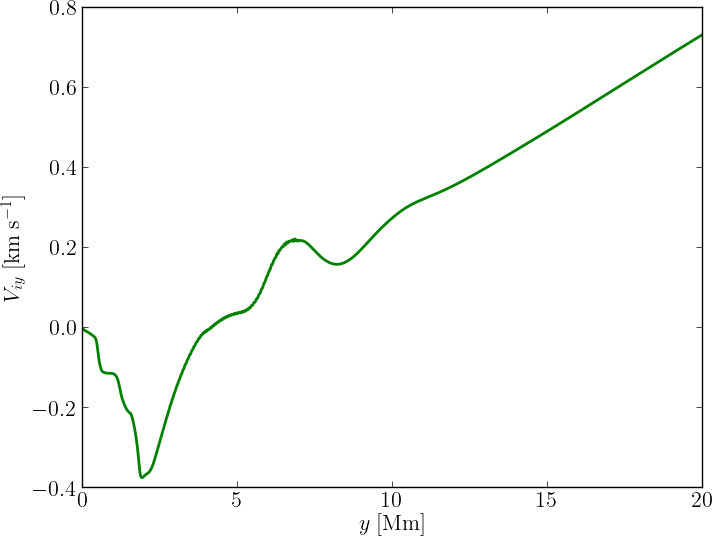}
    \end{center}
    \vspace{-0.2cm}
    \caption{Time--distance plots for $V_{\rm iy}$ (top) and averaged $V_{\rm iy}$ over time  (bottom) for $y_{\rm 0} = 0 \; \rm{Mm}$, $A=0.5 \; \rm{km} \; \rm{s}^{-1}$ (left), and $A= -5 \; \rm{km} \; \rm{s}^{-1}$ (right). The velocity is expressed in units of $\rm{km} \; \rm{s}^{-1}$.}
    \label{fig3}
\end{figure*}
%
Figure 2 illustrates time--distance plots for the ratio between the vertical component of the ion velocity and the local sound speed, $V_{\rm iy}/C_s$, (top) and the relative perturbed temperature of ions, $\delta T_i / T = (T_i - T) / T$, (bottom) in the absence of any initial pulse (i.e.,\ for $A=0$ in Eq.~18).
The signals present in both panels result from numerical errors which are inherent features of every numerical code. As the max~$|V_{\rm iy}/C_s| \approx 0.08$ occurs in the corona, and it indicates that the vertical velocities resulting from the discretization are small in comparison to the local sound speed, we infer that the numerical errors induced in the velocity component $V_{\rm iy}$ are negligible.
In the lower regions of the atmosphere, the numerical errors decrease to max~$|V_{\rm iy}/C_s| \approx 0.02$ in the chromosphere, and max~$|V_{\rm iy}/C_s| \approx 0.0002$ in the photosphere.
Similarly, max~$|\delta T_i / T| \approx 0.065$, showing that the numerical errors detected in the perturbed ion temperature are also negligibly small.
In contrast to the numerical errors in velocity, the errors in the perturbed temperature decrease with height to max~$|\delta T_i / T| \approx 0.03$ in the chromosphere, and max~$|\delta T_i / T| \approx 0.02$ in the corona.
The apparent periodicity  in  $\delta T_i / T$  (Fig.~2, bottom) may  coincide  with  the acoustic cut-off period of the numerical noise.
\subsection{Magnetoacoustic waves}
\subsubsection{Pulse amplitude effects}
We first considered an initial pulse at the bottom of the photosphere ($y_0=0\;$Mm). The initial pulse, given by Eq.~(18), excites magnetoacoustic waves. These waves can get shocked in the chromosphere \citep{Snow_2021}.  The formation of shock waves is clearly seen in the time--distance plots for $V_{\rm iy}$ (Fig.~3, top), and it is stronger for a larger value of $|A|$ (right-top).
These shocks heat plasma through ion--neutral collisions.
Similarly to the gravity action, two-fluid effects introduce dispersion for the ion magnetoacoustic and neutral acoustic waves which are described by Eqs.~$(1) - (13)$.
As a result of this dispersion, the initial pulse spreads into a train of upward- and downward-propagating waves.
The transition region oscillates as a natural consequence of the incoming train of waves which is clearly seen as upward- and downward-propagating waves.
These waves move with the local tube speed which is subsonic and subAlfv\'enic.
As gravity introduces characteristic spatial scales in the form of ion and neutral pressure scale heights, the amplitude of the oscillations in $V_{\rm iy}$ grows with height (top panels).
At the top of the photosphere, namely\ at $\pm 0.5\;$Mm, the signal in $V_{\rm iy}$ is relatively small and reaches a maximum value there of $0.8 \; \rm{km \: s^{-1}}$ for $A=0.5 \; \rm{km \: s}^{-1}$ and $9 \; \rm{km \: s^{-1}}$ for $A=-5 \; \rm{km \: s}^{-1}$. 
Similarly at later times, after $t = 10^3 \; \rm{s,}$ the signal is very weak as the impulsively generated magnetoacoustic waves have already left the system.
Shock waves are generated up to $t \approx 700 \; \rm{s}$.
This means that because of this initial pulse, thermal energy is released until this time because it is mainly generated at the shocks by ion--neutral collisions.
%
%
\begin{figure*}[ht]
    \begin{center}
        \hspace{-0.2cm}
        \includegraphics[width=0.48\textwidth]{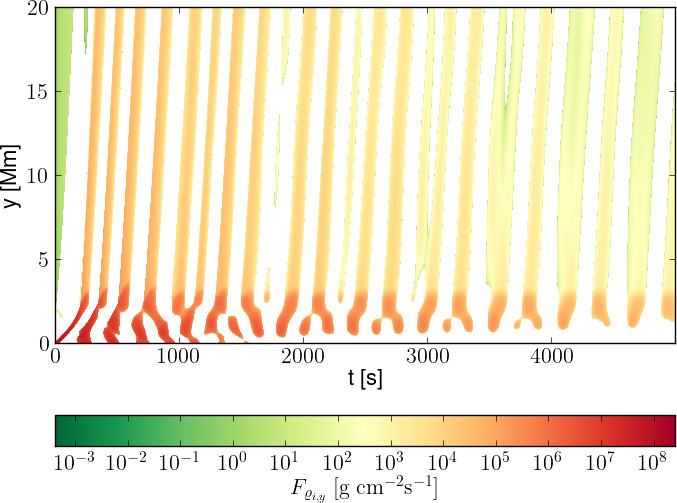}
        \hspace{0.5cm}
        \vspace{0.25cm}
        \includegraphics[width=0.48\textwidth]{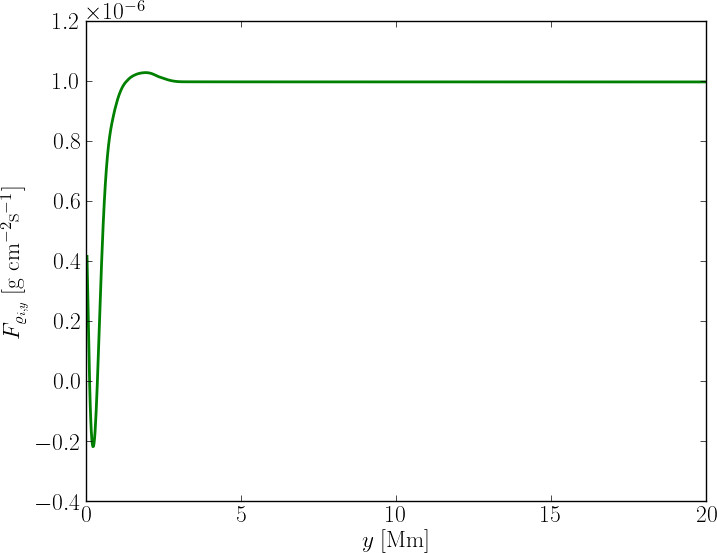}
    \end{center}
    \vspace{-0.5cm}
    \caption{ Time--distance plots for ion mass flux, 
    $F_{\varrho_{i,y}} = \varrho_i V_{\rm iy}$ (left), and its profile  averaged over
time (right) for the initial pulse with width $w = 0.3 \; \rm{Mm}$ 
    and amplitude $A=0.5 \; \rm{km} \: \rm{s}^{-1}$, 
    launched from $y = y_{\rm 0} = 0 \; \rm{Mm}$.}
    \label{fig4_new}
\end{figure*}
%
%
%
%
\begin{figure*}[ht]
    \begin{center}
        \hspace{-0.2cm}
        \includegraphics[width=0.48\textwidth]{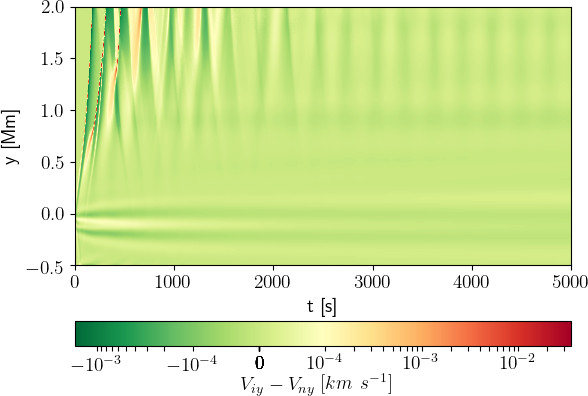}
        \hspace{0.5cm}
        \vspace{0.25cm}
        \includegraphics[width=0.48\textwidth]{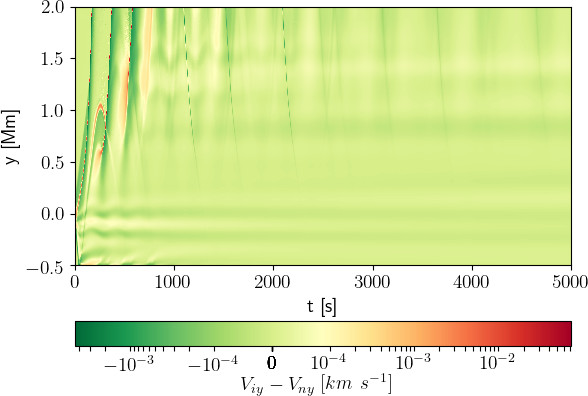}
        \vspace{0.5cm}
        \includegraphics[width=0.48\textwidth]{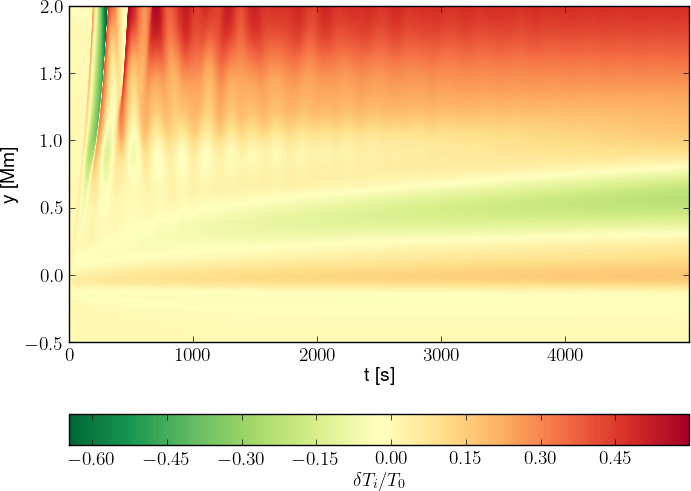}
        \hspace{0.5cm}
        \includegraphics[width=0.48\textwidth]{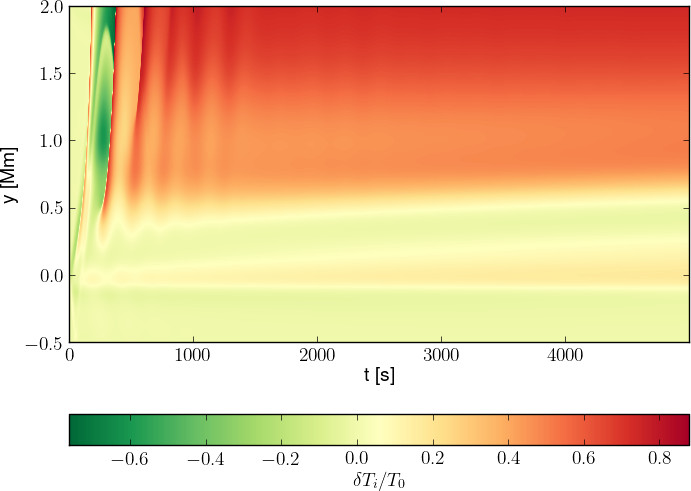}
        \vspace{0.5cm}
        \includegraphics[width=0.48\textwidth]{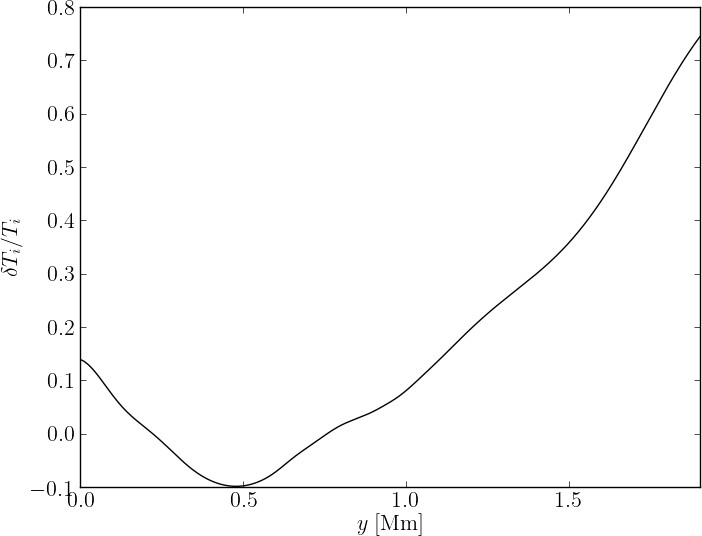}
        \hspace{0.5cm}
        \includegraphics[width=0.48\textwidth]{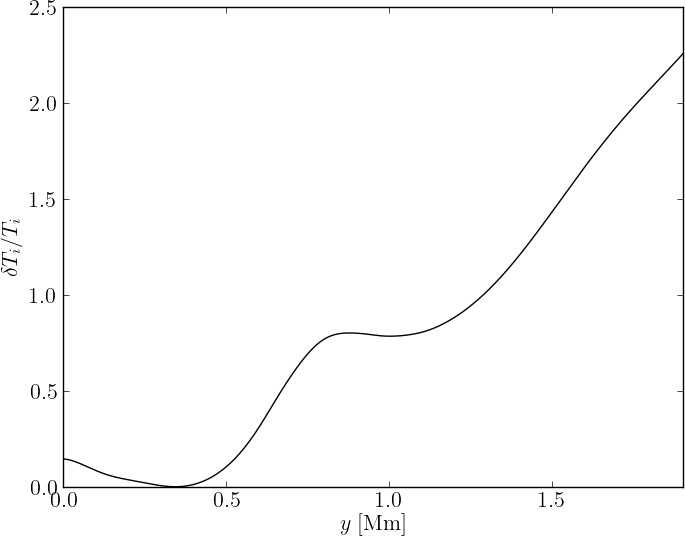}
    \end{center}
    \vspace{-0.5cm}
    \caption{Time--distance plots for the velocity drift, $V_{\rm iy} - V_{ny}$ (top), $\delta T_i/T$ (middle), and averaged $\delta T_i/T$ over time  (bottom) for the pulse with width $w = 0.3 \; \rm{Mm}$, launched from $y = y_{\rm 0} = 0 \; \rm{Mm}$ and with amplitude: $A=0.5 \; \rm{km} \; \rm{s}^{-1}$ (left) and $A= -5 \; \rm{km} \; \rm{s}^{-1}$ (right).}
    \label{fig4}
\end{figure*}
%
%
The initial pulse also induces plasma flows. These are evident from the temporarily averaged vertical ion velocity given as
\begin{equation}
    \langle V_{\rm iy}\rangle  = \frac{1}{t_{\rm 2} - t_{\rm 1}} \int_{t_{\rm 1}}^{t_{\rm 2}} V_{\rm iy} \; dt \, ,
\end{equation}
where $t_{\rm 1} = 0 \; \rm{s}$ and $t_{\rm 2} = 5000 \; \rm{s}$.
The outflows occur in the corona (Fig.~3, bottom).
A higher amplitude pulse results in faster outflows originating from lower levels of the atmosphere.
In particular, for $A = 0.5 \; \rm{km \: s}^{-1}$, $\langle V_{\rm iy}\rangle $ grows almost linearly with height $y$, reaching a value of $0.02 \; \rm{km \; s}^{-1}$ at $y = 20 \: \rm{Mm}$ (Fig.~3, bottom-left).
For $A = -5 \; \rm{km \: s}^{-1}$ the growth of $\langle V_{\rm iy}\rangle $ with $y$ is more abrupt than in the case of $A = 0.5 \; \rm{km \: s}^{-1}$ and $\langle V_{\rm iy}\rangle $ attains a value of about $0.7 \; \rm{km \: s}^{-1}$ at $20 \; \rm{Mm}$.
This speed is smaller than in the semi-empirical data of \cite{AvrettLoeser2008}, where the velocity is of the order of $40 \; \rm{km \: s}^{-1}$ at $y = 3 \; \rm{Mm}$.
Downflows are observed in the lower atmospheric regions.
Such upflows of $\langle V_{\rm iy}\rangle  = 2 \; \rm{km \: s}^{-1}$ and downflows of $\langle V_{\rm iy}\rangle  \textless 10 \; \rm{km \: s}^{-1}$ were reported by \cite{Kayshap_2015} and more recently also by \cite{tian2021upflows}.
%
%
\begin{figure*}[ht]
    \begin{center}
    \hspace{-0.2cm}
    \includegraphics[width=0.48\textwidth]{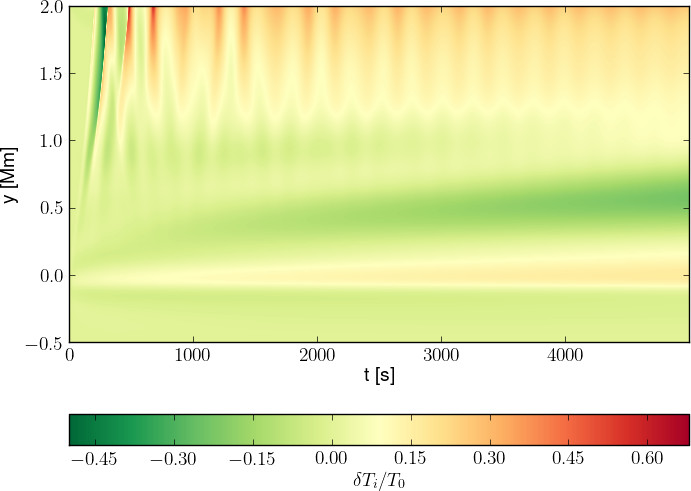}
    \hspace{0.5cm}
    \vspace{0.25cm}
    \includegraphics[width=0.48\textwidth]{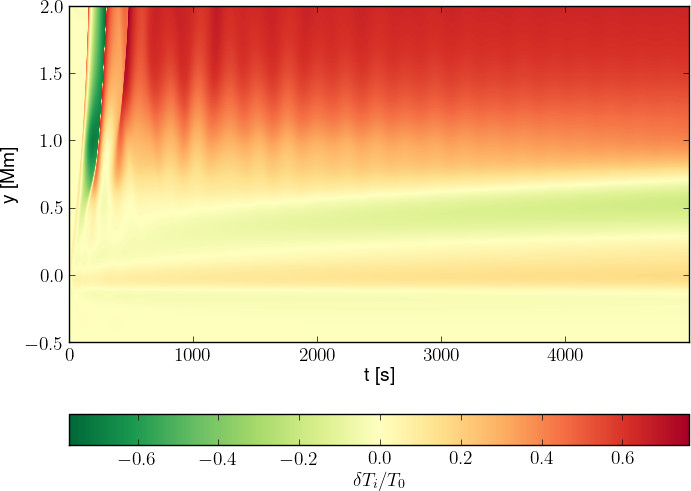}
    \vspace{0.5cm}
    \includegraphics[width=0.48\textwidth]{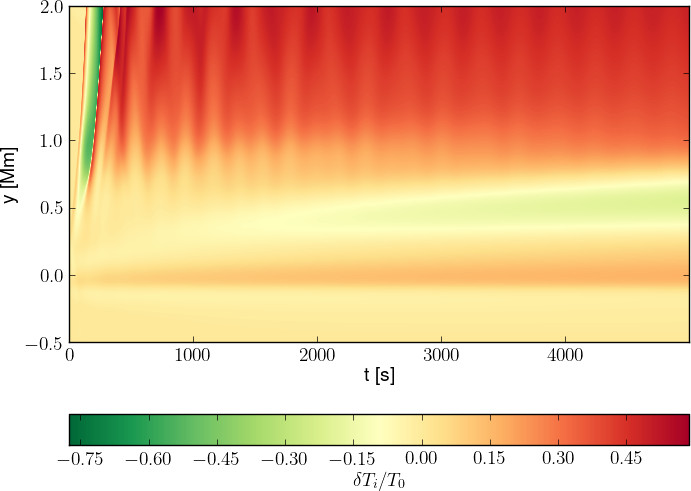}
    \hspace{0.5cm}
    \includegraphics[width=0.48\textwidth]{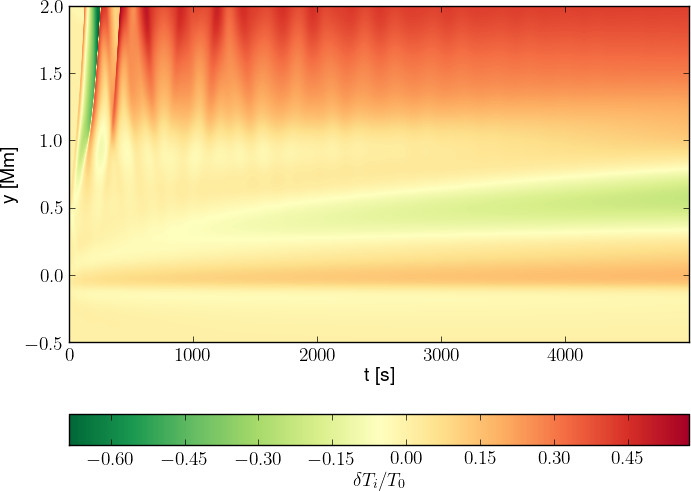}
    \end{center}
    \vspace{-0.5cm}
    \caption{Time--distance plots for $\delta T_i/T$ for the pulse with amplitude $A=2 \; \rm{km} \; \rm{s}^{-1}$ and width $w = 0.1 \; \rm{Mm}$ (top-left) and $w = 0.25 \; \rm{Mm}$  launched from $y = y_{\rm 0} = 0 \; \rm{Mm}$ (top-right), and for $w = 0.3 \; \rm{Mm}$ launched from $y_{\rm 0} = 0.25 \; \rm{Mm}$ (bottom-left) and $y_{\rm 0} = 0.5 \; \rm{Mm}$ (bottom-right).}
    \label{fig5}
\end{figure*}
%

Figure\, \ref{fig4_new} (left panel) illustrates the ion mass flux, 
$F_{\varrho_{iy}}=\varrho_{iy}V_{\rm iy}$, which in contrast to the ion velocity reaches its highest values below the transition region. 
Negative values of the flux are observed only in the initial phase of the wave evolution lasting up to $t \approx 200 \; \rm s$. The magnetoacoustic waves, which are reflected 
from the transition region, are represented by the stripes leaning towards  the $y-$axis.  
The right panel presents the vertical profile of the averaged (over time) ion mass flux which reaches its lowest value of $F_{\varrho_{i,y}} = -0.2 \times 10^{-6} \; \rm g~cm^{-2}s^{-1}$ at $y = 0.2 \; \rm Mm$. Higher up, it abruptly grows  to $F_{\varrho_{i,y}} =10^{-6} \; \rm g~cm^{-2}s^{-1}$ at $y = 1.5 \; \rm Mm$ and then remains essentially constant with height.

Figure~5 (top) illustrates the velocity drift, $\delta V =  V_{\rm iy} - V_{ny}$, which grows with height, reaching its maximum magnitude in the upper chromosphere.
However, at the bottom of the numerical region, ions and neutrals are strongly coupled, and therefore they propagate with almost the same velocity.
As the quantity $\delta V$ attains its largest values in the initial phase, we infer that plasma heating can take place up to $t = 10^3 \; \rm{s}$ after the initial pulse.
The values of the velocity drift are small, which indicates that ions and neutrals are strongly coupled.
In the right-top panel, at the bottom of the photosphere, $y = 0 \; \rm Mm$, and below at $y \approx 0.2 \; \rm Mm$ and $y \approx 0.5 \; \rm Mm$ we observe some oscillatory signatures.
Signatures at $y = 0 \; \rm Mm$, and $y \approx 0.2 \; \rm Mm$ look like quasi-harmonic decaying oscillations while the third one at $y \approx 0.5 \; \rm Mm$ is seen as  faster decaying waves propagating upward.
As they appear at the bottom of the simulation box they, may result from numerical errors.
\\
The middle panels of Fig.~5 show the relative perturbed temperature specified as
\begin{equation}
    \frac{\delta T_i}{T} = \frac{T_i - T}{T} \, .
\end{equation}
This quantity reaches its largest magnitude of $0.55$ at $y \approx 2 \; \rm{Mm}$ for $A= 0.5 \; \rm{km} \; \rm{s}^{-1}$ and $0.8$ at $y \approx 2 \; \rm{Mm}$ for $A= -5 \; \rm{km} \; \rm{s}^{-1}$, while the photosphere is hardly heated at all.
We note that the signals in the top and middle panels exhibit a level of correlation, which means that the chromospheric plasma is heated by the ion--neutral collisions.

From Eqs.~(10) and (11) we infer that the velocity drift, $\delta V$, contributes an important part to the heating and this contribution is particularly effective at shocks, which is clearly  seen in Fig.~5 (top). This was also noticed by \cite{2020arXiv201113469Z} and \cite{Snow_2021} from similar simulations in a different setup. This heating can be quantified by the quantity $\delta T_i / T$ which is displayed on Fig.~5 (middle).

We define the time-averaged relative perturbed temperature as
\begin{equation}
    \left\langle  \frac{\delta T_i}{T} \right\rangle  = \frac{1}{t_{\rm 2} - t_{\rm 1}} \int_{t_{\rm 1}}^{t_{\rm 2}} \frac{\delta T_i}{T} \; dt \, ,
\end{equation}
where $t_{\rm 1} = 0 \; \rm{s}$ and $t_{\rm 2} = 5000 \; \rm{s}$. This quantity is illustrated in the  bottom panel of Fig.~5.
We note that the chromosphere is more strongly heated for higher values of the pulse amplitude $|A|$.
The lowest averaged temperature is seen in the top of the photosphere and above this region the atmosphere is heated to a greater temperature.
For $A = -5 \; \rm{km} \; \rm{s}^{-1}$, all values of $\left\langle  \delta T_i/T \right\rangle $ are positive and the minimum of $\left\langle  \delta T_i / T \right\rangle  \approx 0$ at $y \approx 0.4 \; \rm{Mm}$ takes place at the equilibrium temperature minimum \citep{AvrettLoeser2008}, where the magnetoacoustic waves experience reflections. $\left\langle  \delta
T_i/T \right\rangle $ reaches a value of about $2.3$ at $y = 2 \; \rm{Mm}$, at the top of the chromosphere. A characteristic flattening is present within the lower levels of the middle chromosphere, from $y \approx 0.8 \; \rm{Mm}$ to $y \approx 1 \; \rm{Mm}$.
Above this plateau, $\left\langle  \delta T_i/T \right\rangle $ suddenly increases with $y$ (bottom-right).

We note that the temperature is reduced permanently in the photosphere with essentially unseen waves in $\delta T_i/T$ (Fig.~5, middle).
In the chromosphere, we observe oblique stripes which are leaned towards the vertical axis and represent magnetoacoustic waves.
In the case of $A=0.5 \; \rm{km} \; \rm{s}^{-1}$, max $(\delta T_i / T) \approx 0.6$  at $y = 1.8 \; \rm{Mm}$, which is located in the chromosphere (middle-left).
As for larger absolute values of amplitude, for example\ for $A = - 5 \; \rm{km \: s^{-1}}$ (middle-right), the generated shocks are more pronounced than for $A=0.5 \; \rm{km}\: \rm{s}^{-1}$, and result in more intensive heating with max$\;(\delta T_i / T) \approx 0.8$ in the middle of the chromosphere at $y = 1.1 \; \rm{Mm}$ (Fig.~5, middle).
\subsubsection{Pulse width effect}
In order to study how the width of the pulse, $w$, affects the plasma heating, we performed some parametric studies.
From Fig.~6 (top) we infer that increasing $w$ from $0.1 \; \rm{Mm}$ to $0.25 \; \rm{Mm}$ results in a significant heating of the chromosphere.
However, for $w = 0.1 \; \rm{Mm,}$ only the top layer of the chromosphere is heated (top-left), while a wider pulse width, $w = 0.3 \; \rm{Mm}$, results in heating of the whole chromosphere (top-right). 
Similarly, the cooling during the first $200\;$s is more pronounced for the wider pulse, and results from the plasma being rarefied.
\subsubsection{Pulse launching height effect}
Figure~6 (bottom) shows time--distance plots for $\delta T_i/T$ for different pulse-launching heights, that is,\ for different values of the parameter $y_{\rm 0}$.
A higher value of $y_{\rm 0}$ results in a smaller value of max $(\delta T_i/T)$ at a given height $y$.
This is the consequence of a pressure scale-height over which, according to linear theory, the amplitude of the signal grows $e-$times \citep [e.g.,] [] {2005LRSP23N}.
For the pulse launched from a higher level $y_{\rm 0}$, with the same amplitude, the signal passes fewer scale heights, reaching a given level $y$.
Consequently, $\delta V$ and $\delta T_i/T$ are smaller (at a given height $y$) for a larger value of $y_{\rm 0}$.

Figure~7 illustrates the relative perturbed temperature averaged over time and height, $H$, defined as
\begin{equation}
    H = \frac{1}{y_{\rm 1}-y_{\rm 0}} \int_{y_{\rm 0}}^{y_{\rm 1}} \left\langle  \frac{\delta T_i}{T} \right\rangle  \; dy \, .
\end{equation}
Here, $y_{\rm 1} = 1.9 \; \rm{Mm}$.
%
\begin{figure}[!ht]
    \begin{center}
    \hspace{-0.2cm}
    \includegraphics[width=0.48\textwidth]{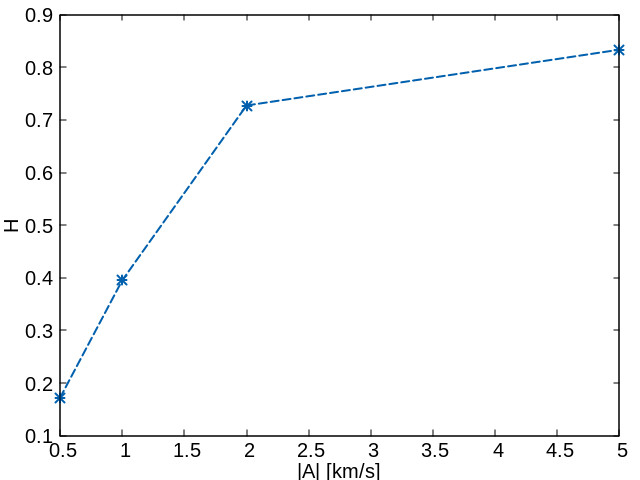}
    \vspace{0.4cm}
    \includegraphics[width=0.48\textwidth]{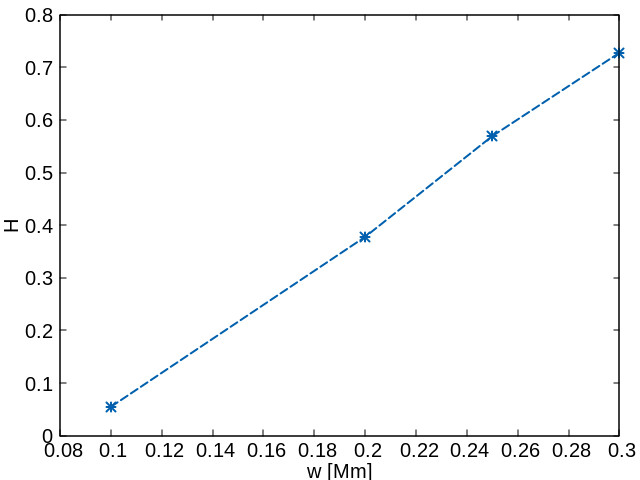}
    \vspace{0.5cm}
    \includegraphics[width=0.48\textwidth]{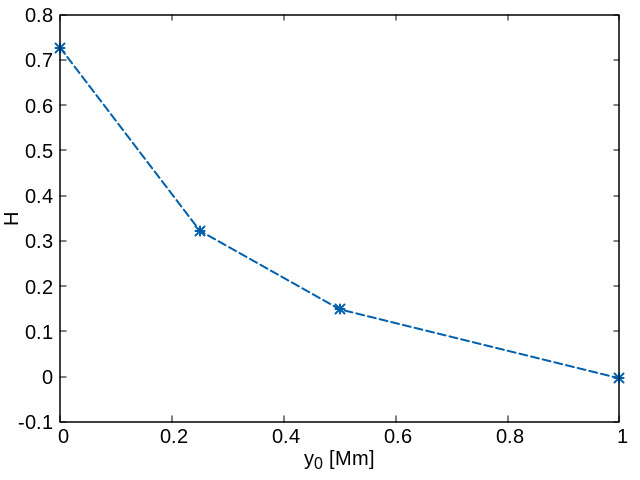}
     \end{center}
    \vspace{-0.5cm}
    \caption{Relative perturbed temperature averaged over time and height, $H$, vs: $|A|$, for $y_{\rm 0} = 0 \; \rm{Mm}$ and $w = 0.3 \; \rm{Mm}$ (top); $w$ for $A= 2 \; \rm{km \: s}^{-1}$ and $y_{\rm 0} = 0 \; \rm{Mm}$ (middle); and $y_{\rm 0}$ for $A= 2 \; \rm{km \: s}^{-1}$ and $w = 0.3 \; \rm{Mm}$ (bottom). }
    \label{fig6}
\end{figure}
%
%
%
\begin{figure}[!ht]
    \begin{center}
    \hspace{-0.2cm}
    \includegraphics[width=0.48\textwidth]{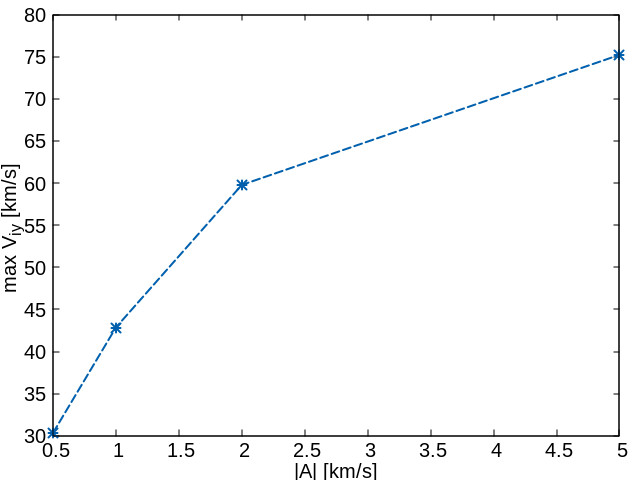}
    \vspace{0.5cm}
    \includegraphics[width=0.48\textwidth]{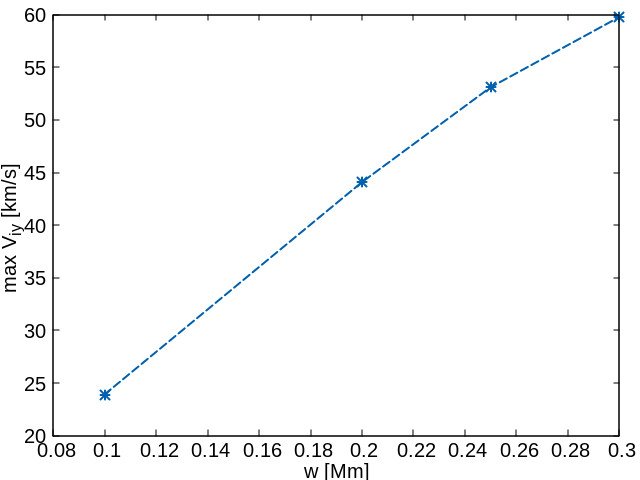}
    \vspace{0.5cm}
    \includegraphics[width=0.48\textwidth]{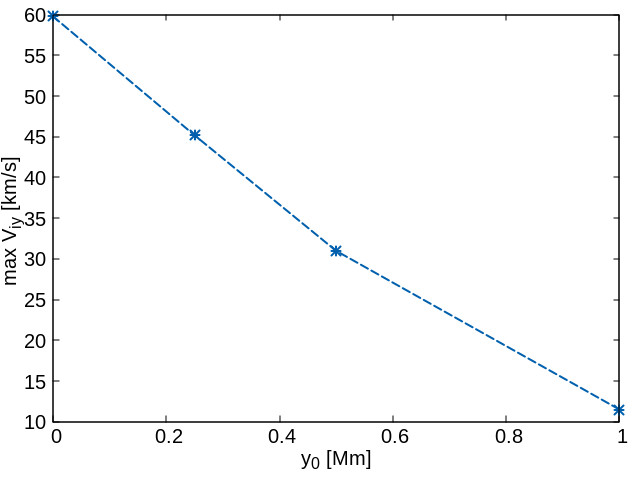}
     \end{center}
    \vspace{-0.5cm}
    \caption{Maximum value of $V_{\rm iy}$ vs: $|A|$, for $y_{\rm 0} = 0 \; \rm{Mm}$ and $w = 0.3 \; \rm{Mm}$ (top); $w$ for $A= 2 \; \rm{km \: s}^{-1}$ and $y_{\rm 0} = 0 \; \rm{Mm}$ (middle); and $y_{\rm 0}$ for $A= 2 \; \rm{km \: s}^{-1}$ and $w = 0.3 \; \rm{Mm}$ (bottom). }
    \label{fig7}
\end{figure}
%
%
%
\begin{figure}[!ht]
    \begin{center}
    \hspace{-0.2cm}
    \includegraphics[width=0.48\textwidth]{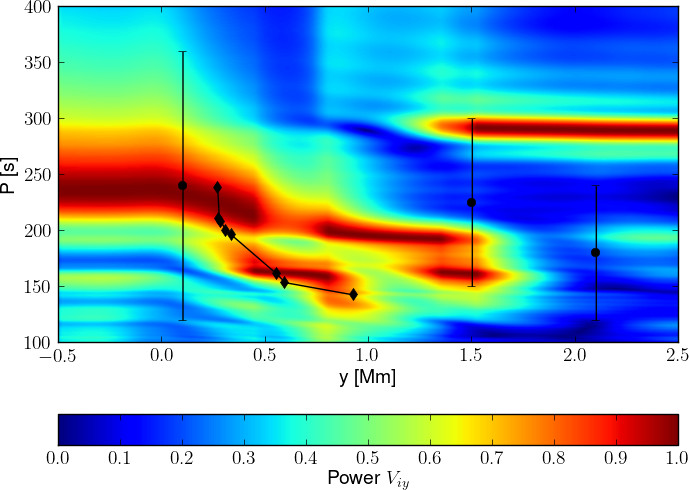}
    \hspace{0.5cm}
    \includegraphics[width=0.48\textwidth]{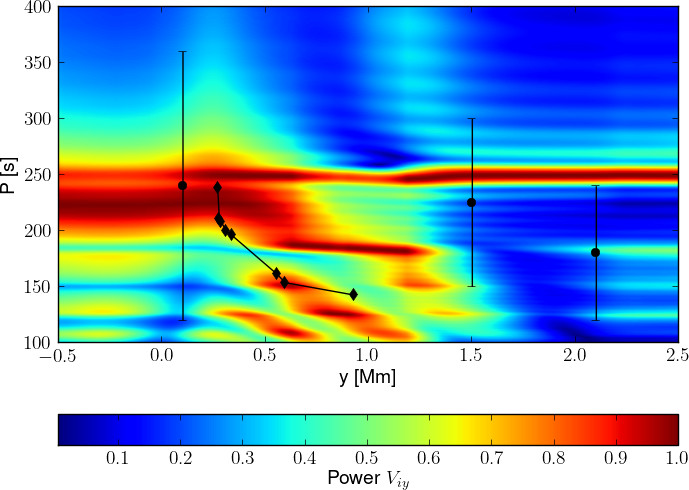}
     \end{center}
    \vspace{-0.25cm}
    \caption{Fourier period $P$ for $V_{\rm iy}$ vs. height, in the case of the initial pulse with amplitude $A=0.5 \; \rm{km} \; \rm{s}^{-1}$, width $w = 0.3 \; \rm{Mm}$, and launched from $y = y_{\rm 0} = 0 \; \rm{Mm}$ (top) and $y = y_{\rm 0} = 0.25 \; \rm{Mm}$ (bottom). The observational data of \cite{Wi_niewska_2016} and \cite{101093mnrassty1861} are represented by diamonds and dots with the vertical error bars, respectively.}
    \label{fig:my_label}
\end{figure}
%
%
The higher value of the pulse amplitude $|A|$ results in more significant heating of the photosphere and chromosphere (top). 
Similarly, a higher magnitude of the pulse width $w$ leads to a higher value of $H$  (middle). 
Finally, increasing the pulse launching height $y_{\rm 0}$ corresponds to less heating below the transition region with a fall-off of $H(y_{\rm 0})$ resulting from the pressure scale height (bottom). 
For a larger value of $y_{\rm 0}$, the distance to the transition 
region covers fewer pressure scale heights and the signal has less chance 
to increase its amplitude and consequently to generate stronger shocks 
with less thermal energy release. 
The value of $H = 1 $ would correspond to an average relative temperature increase in the photosphere and chromosphere by $100\%$.
However, we do not reach such a value in our results because the amplitude of the pulse is insufficient to induce such strong heating.
%
%
\subsubsection{Outflow velocities and Fourier power spectrum}
Figure~8 shows the maximum ion outflow velocity, $\rm{max}(V_{\rm iy})$, for different pulse parameters.
Similarly to in Fig.~7, a pulse with larger absolute amplitude (top) and width (middle) results in a higher value of the maximum outflow velocity $\rm{max}(V_{\rm iy})$.
As we increase the launching altitude of the initial pulse, $y_{\rm 0}$, the velocity naturally decreases (bottom).

Figure~9 presents Fourier power spectrum for wave period $P$ versus height.
The initial pulse has a Gaussian spectrum of wave number $k$ which results in a spectrum of period $P$.
The steepening of the magnetoacoustic waves results from the growing wave amplitude with height. Hence, waves with shorter wavelengths and wave periods are present for higher $y$-values in their Fourier spectra.
For the pulse launched from $y=y_{\rm 0}=0 \; \rm{Mm}$ (top), the main wave period of the downward-propagating waves becomes approximately the same for all values of $y<0 \; \rm{Mm}$ and is equal to about $250\;$s.
Higher up, however, the period $P$ decays with increasing $y$, and attains values close to $200 \; \rm{s}$.
As a result of the cut-off only short-period waves can propagate upwards while long-period waves become evanescent.
Hence, the relative contribution of long $P$ waves weakens with increasing $y$.
We note that some of our data fits the observational findings of \cite{Wi_niewska_2016}, represented by diamonds over-plotted on the power spectra, and \cite{101093mnrassty1861}, denoted by dots.
The agreement of the theory with the observational data indicates that the results can be used to determine the background structure of the solar atmosphere and confirms that wave generation by the solar granulation in the partially ionized plasma dominates the behavior of the waves.
For a pulse at the bottom of the photosphere ($y=y_{\rm 0}=0 \; \rm{Mm}$, Fig.~9 (top)), a jump in the dominant wave period is observed close to the height $y = 1.5 \; \rm{Mm}$, which reaches a magnitude of $300 \; \rm{s}$.
The signal launched from the bottom of the photosphere with the main period $ P = 250 \; \rm{s}$ thus reaches the corona with $P = 300 \; \rm{s}$.
The wave periods in the photosphere are lower than $P_{\rm ac}$ in this region, because max $P_{\rm ac} = 240 \; \rm{s}$ at $y = 0.5 \; \rm{Mm}$, which means that magnetoacoustic waves are evanescent.
However, in the corona above $y = 2.3 \; \rm{Mm,}$ the acoustic cut-off period reaches values larger than $P = 300 \; \rm{s}$.
However, this is not the case for a pulse launched at a somewhat greater height, in the middle of the photosphere, namely\ for $y=y_{\rm 0}=0.25 \; \rm{Mm}$ (see Fig.~9 (bottom)).
Moreover, in this case the layers below the photosphere ($y<0$) oscillate with a dominant wave period of $225\;$s. In the photosphere, this wave period is also dominant and this time it increases slightly with height in the photosphere. Moreover, in this case there is a second dominant period, namely $250\;$s. In fact, this wave period is also the dominant one in the upper chromosphere and low corona in this case. 
%
\section{Summary and Conclusions}
In this paper we consider two-fluid ion magnetoacoustic and neutral acoustic waves that are initially (at $t = 0 \; \rm{s}$)  excited by a pulse in a vertical component of the ion and neutral velocities.
The full set of two-fluid equations is solved by the JOANNA code \citep{Wojciketal2019a}.
The results of the numerical simulations presented in this paper can be summarized as follows.

The triggered magnetoacoustic waves transform into shock waves and through ion--neutral collisions they convert their energies into thermal energy contributing to heating of the chromospheric plasma.
Increasing the amplitude of the pulse leads to a significant heating of the chromosphere, which for the highest absolute value of the amplitude considered, $A = - 5 \; \rm{km} \; \rm{s}^{-1}$, is clearly seen to occur in the upper part of the chromosphere, above the  level $y = 1 \; \rm{Mm}$.
Moreover, the heating of the chromosphere is correlated with a velocity drift which increases with height up to $y = 2 \; \rm{Mm}$.
This means that the chromosphere is heated by the ion--neutral collisions.
The parametric studies we performed for various widths of the pulse show that a wider pulse corresponds to more plasma heating spread over the whole chromosphere.
On the other hand, the investigation of quasi-periodic MHD waves by \cite{2005SSRv..121..115N} showed that a wider pulse is less effective in generating quasi-periodic wave trains.
The problem of impulsively generated wave trains was also studied by \cite{2017ApJ...836....1Y} who concluded that the diversity of group speed characteristics has an impact on the temporal evolution of impulsively generated wave trains.
A pulse launched from a higher level (with the same amplitude), on the other hand, results in less heating as the amplitude of upwardly propagating signal cannot grow significantly with height before reaching the transition region.
Both the plasma temperature and maximum ion velocity increase with increasing pulse amplitude and width, while the opposite trend is seen for higher values of $y_{\rm 0}$, that is,\ for launching the pulse higher up.
The pulses considered in this paper result in shocks that lead to a release of thermal energy which in turn helps to sustain the chromosphere.
We arrive at a further conclusion, namely that even a single pulse can drive vertical plasma outflows which higher up can result in (contributions to) the origin of the solar wind.
%
\section*{Acknowledgments}
The authors express their gratitude to the reviewer for his/her 
stimulating comments and suggestions.
The JOANNA code was written by Dr.\ Dariusz W\'ojcik. This work was done within the framework of the project from the National Science Centre (NCN) grant nos. 2017/25/B/ST9/00506 and 2020/37/B/ST9/00184.
Numerical simulations were performed on the LUNAR cluster at the Institute of Mathematics of University of M. Curie-Skłodowska, Lublin, Poland.
SP is supported by the project EUHFORIA 2.0 which has received funding from the European Union’s Horizon 2020 research and innovation programme under grant agreement No 870405, and by the projects C14/19/089  (C1 project Internal Funds KU Leuven), G.0D07.19N  (FWO-Vlaanderen), SIDC Data Exploitation (ESA Prodex-12), and the Belspo projects BR/165/A2/CCSOM and B2/191/P1/SWiM.

\bibliographystyle{aa} 
\bibliography{main.bbl}
\end{document}